\documentclass[fleqn,10pt,a4paper]{wlscirep}
\usepackage[utf8]{inputenc}
\usepackage[T1]{fontenc}
\usepackage{dsfont}
\usepackage{bm}

\title{Experimental realization of Fermi-Pasta-Ulam-Tsingou recurrence in a long-haul optical fiber transmission system}

\author[1,2]{Jan-Willem Goossens}
\author[1,*]{Hartmut Hafermann}
\author[2]{Yves Jaou\"en}
\affil[1]{Optical Communication Technology Lab, Paris Research Center, Huawei Technologies France,\newline 92100 Boulogne-Billancourt, France}
\affil[2]{LTCI, T\'el\'ecom Paris, Universit\'e Paris-Saclay, 91120 Palaiseau, France}

\affil[*]{hartmut.hafermann@huawei.com}


\begin{abstract}
The integrable nonlinear Schr\"odinger equation (NLSE) is a fundamental model of nonlinear science which also has important consequences in engineering. The powerful framework of the periodic inverse scattering transform (IST) 
provides a description of the nonlinear phenomena modulational instability and Fermi-Pasta-Ulam-Tsingou (FPUT) recurrence in terms of exact solutions. It associates the complex nonlinear dynamics with invariant nonlinear spectral degrees of freedom that may be used to encode information.
While optical fiber is an ideal testing ground of its predictions, 
maintaining integrability over sufficiently long distances to observe recurrence, as well as synthesizing and measuring the field in both amplitude and phase on the picosecond timescales of typical experiments is challenging.
Here we report on the experimental realization of FPUT recurrence in terms of an exact space-time-periodic solution of the integrable NLSE in a testbed for optical communication experiments. 
The complex-valued initial condition is constructed by means of the finite-gap integration method, modulated onto the optical carrier driven by an arbitrary waveform generator and launched into a recirculating fiber loop with periodic amplification. 
The measurement with an intradyne coherent receiver after a predetermined number of revolutions provides a non-invasive full-field characterization of the space-time dynamics.
The recurrent space-time evolution is in close agreement with theoretical predictions over a distance of 9000 km. Nonlinear spectral analysis reveals an invariant nonlinear spectrum. The space-time scale exceeds that of previous experiments on FPUT recurrence in fiber by three orders of magnitude.
\end{abstract}
\begin{document}

\flushbottom
\maketitle

\thispagestyle{empty}

The integrable NLSE is an important exactly solvable model for the study of nonlinear phenomena. An example is modulational instability (MI)~\cite{Agrawal2000}, an exponential amplification of periodic random fluctuations at the expense of a pump wave that has been suggested as a possible mechanism for the generation of rogue waves~\cite{Dudley2014}. The reversal of this process can give rise to repeated cycles of growth and decay, which constitute a realization of FPUT recurrence~\cite{Fermi1955,Simaeys2001}. In the framework of the IST, these phenomena find a description in terms of exact solutions~\cite{Tracy1984,Tracy1988} associated with conserved nonlinear spectral degrees of freedom. From an engineering perspective, the prospect of encoding information in the invariant nonlinear spectrum is of high interest for optical communication systems, which today are limited by nonlinear interference~\cite{Turitsyn2017}.

Various predictions of the underlying analytical NLSE theory have been observed in optical fiber experiments, including solitons~\cite{Mollenauer1980}, Akhmediev breathers~\cite{Dudley2009} and their collisions~\cite{Frisquet2013}, the Peregrine soliton~\cite{Kibler2010}, and the Kuznetsov-Ma soliton~\cite{Kibler2012}. Such experiments are not without challenges. They are typically  conducted at average signal powers up to a few Watts~\cite{Erkintalo2011,Tikan2018,Frisquet2013,Kibler2012,Xu2019,Mussot2018,Naveau2019}. The dynamics take place over distances of several hundred meters up to few kilometers and on the picosecond scale.
At such timescales, the generation of arbitrary initial conditions is difficult and has been approximated by amplitude modulation based on dual-frequency excitation~\cite{Dudley2001, Dudley2009}, by beating of two narrow-linewidth lasers to create a low-frequency modulation~\cite{Kibler2010}, or excitation of superpositions of complex exponentials with tuned relative phases and amplitudes. The latter can be obtained from an optical frequency comb shaped with a programmable optical filter~\cite{Frisquet2013}.
The observation of the spatial dynamics has been achieved with fiber cut-back experiments~\cite{Kibler2012}. Simultaneous observation of amplitude and phase information can be realized by frequency-resolved optical gating (FROG)~\cite{Kibler2010} or nonlinear digital holography~\cite{Tikan2018}. 

Another challenge is to approximate the ideal integrable system over sufficiently long distances. 
While a recirculating fiber loop with repeated Raman amplification~\cite{Kraych2019,Kraych2019b} can extend the reach substantially, it has been shown that deviations from predictions of the integrable system due to a small residual damping can be significant~\cite{Kraych2019}.
Recently a non-destructive full-field characterization of the nonlinear dynamics along an optical fiber based on a heterodyne optical time-domain reflectometer (HOTDR) revealed the broken symmetry of FPUT recurrence through the study of space-time-periodic structures~\cite{Mussot2018,Naveau2019}. However the analysis has been limited to a small number of spectral components of the backscattered light.

Coherent optical transmission systems, on the other hand, transmit over vastly different scales and routinely encode and detect information in both amplitude and phase. Recirculating fiber loops are widely used to conveniently simulate transmission over transcontinental distances in the laboratory~\cite{Bergano1995}. Time dynamics typically occur on the scale of nanoseconds. In experimental testbeds, optical modulators driven by specifically designed electrical signals give precise control over the optical waveform in both amplitude and phase, which provides a resolution on the order of about ten picoseconds. Coherent receivers allow independent detection of the real and imaginary part of the electrical field which are sampled via a real-time oscilloscope at sampling rates comparable to that of the signal generation.

In this paper, we exploit these properties to achieve an experimental realization of integrable NLSE dynamics: The propagation through the recirculating fiber loop with periodic Erbium-doped fiber amplification (EDFA) closely approximates an \emph{effective} integrable model, provided that the span length is sufficiently short compared to the scale of the spatial evolution of the waveform.
The setup achieves a non-invasive full-field characterization with a spatial resolution determined by the recirculation loop length and it allows us to target a desired initial condition in amplitude and phase with a high sampling rate. To exploit this, we employ the so-called finite-gap integration theory to obtain an initial condition that provably corresponds to a solution that is (quasi-)periodic not only in time, but also in space, as a clear hallmark of FPUT recurrence.
By carefully designing the periodic waveform to match the experimental scales, the setup allows us to observe the FPUT recurrent behavior beyond a complete growth and decay cycle and over thousands of kilometers, differing in scale of space, time, and power by three orders of magnitude compared to previous experiments observing FPUT recurrence of the NLSE.
Finally, by employing the amplitude and phase information to perform \emph{nonlinear spectral analysis} via the IST, we  observe a well-defined conserved nonlinear spectrum in close agreement to the predictions of the finite-gap theory.

\section*{Results}

\subsection*{Finite-gap solutions}

Finite-gap solutions arise in the algebro-geometric treatment of certain integrable equations under  periodic boundary conditions~\cite{Kotlyarov1976,Its1976,Ma1981}. They establish a deep mathematical link between the NLSE and the theory of compact Riemann surfaces and algebraic geometry~\cite{Belokolos1994} and had important influence on mathematical physics since their discovery more than 40 years ago~\cite{Matveev2007}.
Finite-gap solutions derive their name from the structure of their underlying nonlinear spectrum. They are defined through the IST in terms of a \emph{finite} number of nonlinear spectral parameters. An important characteristic is that the main IST spectrum remains invariant under propagation of the solution governed by the NLSE -- a property which bears the potential to eliminate nonlinear interference limiting data rates in modern fiber optic communication systems~\cite{Turitsyn2017}.

\begin{figure}[t]
\center{{\bf a}\includegraphics[width=0.3\textwidth]{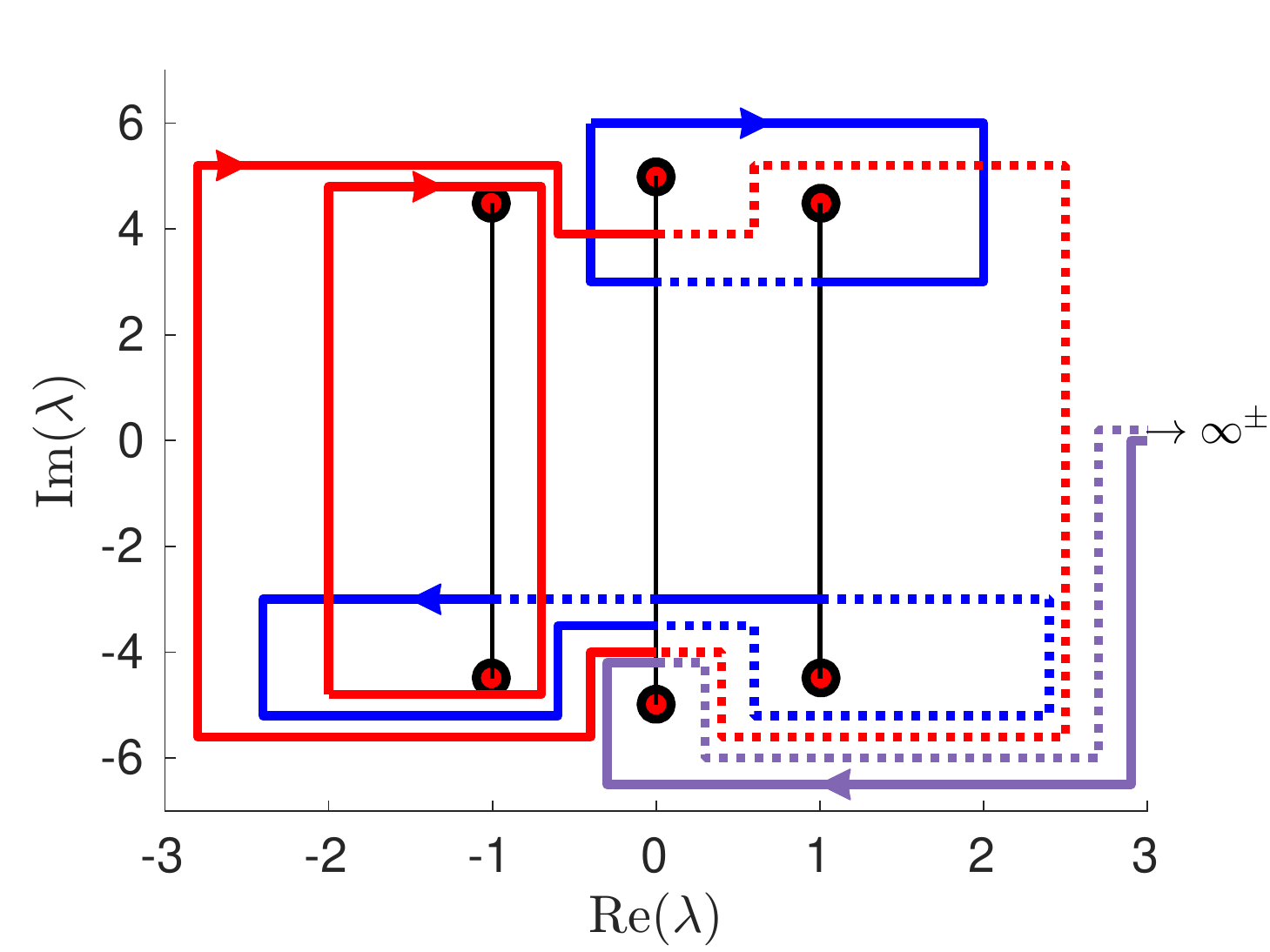}\hspace{10em}{\bf b}\includegraphics[width=0.25\textwidth]{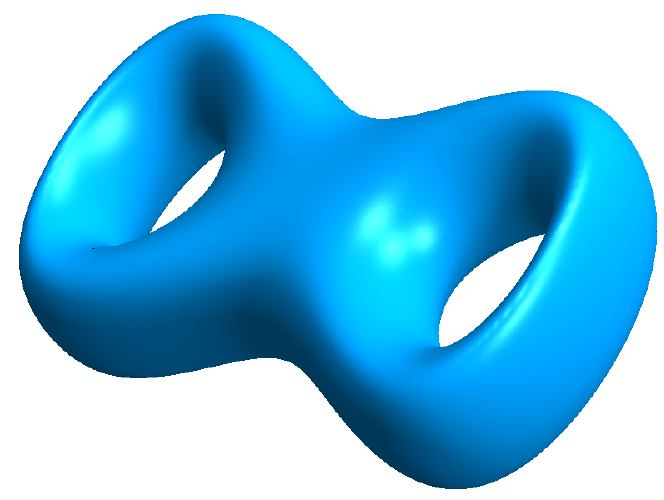}}
\caption{\label{fig:cycles} {\bf Nonlinear spectrum and topology of the hyperelliptic Riemann surface.} {\bf a}, Nonlinear spectrum of the finite-gap solution (red dots) of the experiment and the associated branch cuts (black lines). Lines with arrows are projections of the $a$-cycles (red), and $b$-cycles (blue).
The lines change between solid and dotted when sheet changes occur at the branch cuts.
Another path extends to the points $\infty^\pm$ at infinity on the two sheets of the Riemann surface. 
{\bf b}, Topology of the hyperelliptic Riemann surface of the genus-2 solution. In general, a genus-$g$ solution is topologically equivalent to a torus with $g$ holes. 
}
\end{figure}

Finite-gap solutions have provided analytical insight into the study of MI~
\cite{Tracy1988}. Here we construct solutions of the NLSE which exhibit periodic growth-decay cycles, as a hallmark of FPUT recurrence. 
In its dimensionless form, the focusing NLSE reads
\begin{equation}
i\frac{\partial \psi(\zeta,\tau)}{\partial \zeta} + \frac{\partial^2 \psi(\zeta,\tau)}{\partial \tau^2} + 2|\psi(\zeta,\tau)|^2 \psi(\zeta,\tau) = 0,
\label{eq:nlse}
\end{equation}
where $\zeta$, $\tau$ are normalized space and time coordinates.
In the IST, the field $\psi(\zeta,\tau)$ plays the role of a scattering potential. 
For periodic potentials, the main IST spectrum is a discrete set of  points in the complex plane for which the Bloch functions of the scattering problem are (anti-)periodic.
A solution $\psi(\zeta,\tau)$ is called finite-gap or finite-band  if it can be described in terms of a \emph{finite} number of spectral components. 	
The main spectrum consists of $N$ complex-conjugate pairs $\lambda_j$, $\bar{\lambda}_j$, $j=1\ldots N$, and remains invariant when $\psi(\zeta,\tau)$ evolves according to the NLSE.
The dynamics are encoded in the $N$ auxiliary spectrum variables $\mu_j(\zeta,\tau)$.
They evolve on the two-sheeted Riemann surface of genus $g=N-1$ of an algebraic curve~\cite{Its1976,Belokolos1994,Tracy1984}  
$$
\Gamma = \{(w,\lambda) : w^2 = P(\lambda)\},\quad P(\lambda)=\prod_{j=1}^{g+1} (\lambda-\lambda_j)(\lambda-\bar{\lambda}_j),\quad \rm{Im}\, \lambda_j\neq 0.
$$
The solution of the Jacobi inversion problem on the hyperelliptic  Riemann surface leads to exact solutions in terms of $g$-dimensional Riemann $\theta$-functions in the form
\begin{equation}
\psi(\zeta,\tau) = K\frac{\theta\left(\frac{1}{2\pi}(\underline{k}\zeta+\underline{\omega}\tau+\underline{\delta}^+)|\bm{\tau}\right)}{\theta\left(\frac{1}{2\pi}(\underline{k}\zeta+\underline{\omega}\tau+\underline{\delta}^-)|\bm{\tau}\right)} e^{ik_0\zeta+i\omega_0 \tau}, \quad \text{where}\quad
\theta(\underline{x}|\bm{\tau}) = \sum_{m_1=-\infty}^\infty \ldots\sum_{m_g=-\infty}^\infty \exp(\pi i\underline{m}\bm{\tau}\underline{m}+ 2\pi i\underline{m}\cdot\underline{x}).
\label{eq:thetasol}
\end{equation}

For each value of $\lambda$ there are two values of $w=\pm\sqrt{P(\lambda)}$. Fig.~\ref{fig:cycles} {\bf a} shows the branch cuts between pairs of eigenvalues $\lambda_j,\bar{\lambda_j}$ for $g=2$ where the function $w(\lambda)=\sqrt{P(\lambda)}$ changes sign. To make $w(\lambda)$ single-valued, consider two sheets of the same complex plane on which it takes a definite value. The projective plane $S=\mathds{C}\cup\{\infty\}$ is topologically equivalent to a sphere. If we cut the spheres and glue them together at their three branch cuts we obtain two spheres connected by three tubes, or equivalently, a torus with two holes: a genus-2 surface (Fig.~\ref{fig:cycles} {\bf b}).
There are $2g$ distinct closed curves, which can neither be deformed continuously into each other, nor shrunk to zero. 
We can imagine the $g$ $a$-cycles going around the handles in Fig.~\ref{fig:cycles} {\bf b} the short way, while the $b$-cycles go around the long way. 
They give rise to the $g$ individual phases in the Riemann $\theta$-functions.
Most of the parameters in Eq.~\ref{eq:thetasol}, in particular the $g$-dimensional vectors $\underline{k}$ and $\underline{\omega}$  which determine the space and time period of the solution, are given by integrals over holomorphic differentials over the basis of cycles (see Methods and Fig.~\ref{fig:cycles} {\bf a}).

Finite-gap solutions can hence be characterized by the genus of their underlying Riemann surface.
Genus-0 solutions are plane waves, while genus-1 solutions are soliton-like and preserve the shape of the amplitude. Waveforms with the simplest nonlinear spectrum that exhibit nontrivial space-time dependence are genus-2. 
The Akhmediev breather, the Kuznetsov-Ma soliton and the Peregrine soliton are in fact limits of periodic genus-2 solutions which are known as solitons on finite background (SFB), and for which the period goes to infinity in space, time, or both~\cite{Randoux2016}.

\begin{figure}[t]
\center{\includegraphics[width=0.9\textwidth]{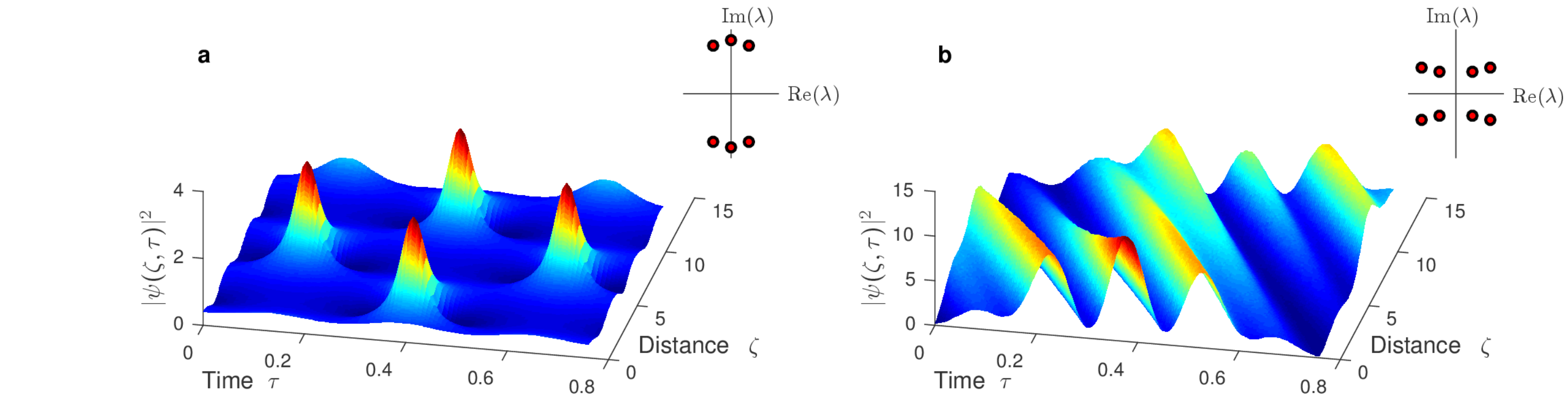}}
\caption{\label{fig:solutions} {\bf Exact solutions of the NLSE and schematic of nonlinear spectrum.} {\bf a}, Genus-2 solution with main spectrum $\{\lambda_j\} = \{-1\pm 4.5i, \pm 5i, 1\pm 4.5i\}$. The solution is exactly periodic in time and quasi-periodic in space.
{\bf b}, Genus-3 solution with main spectrum $\{\lambda_j\} = \{-11.5\pm 5i, -10.5\pm 4i, 10.5\pm 4i, 11.5\pm 5i\}$.
}
\end{figure}

Figure~\ref{fig:solutions} shows two examples of finite-gap solutions. 
The genus-2 solution in {\bf a} develops high peaks surrounded by troughs of minimal intensity (dark blue regions) starting from an almost flat initial condition and clearly shows out-of-phase recurrent behavior~\cite{Mussot2018}. It has the symmetric spectrum $\{-a\pm bi, \pm ci, a\pm bi\}$, which corresponds to an algebraic curve with two involutions (complex conjugation and negation) and implies the existence of double coverings of two genus-1 surfaces by the genus-2 surface. This property  
allows the linear combination of the cycles on the genus-2 surface in terms of cycles on two genus-1 surfaces with integer coefficients. As a consequence, wavevectors and frequencies become integer multiples of each other ($k_1=2k_2$, $\omega_1=0$, $\omega_2=\omega$) so that the solution is exactly periodic in time ($\omega_0=0$) and quasi-periodic (periodic up to a phase) in space~\cite{Smirnov2013}. 
The argument $\underline{\omega}\tau+\underline{k}\zeta$ in Eq.~\ref{eq:thetasol} remains invariant under a simultaneous shift of half a period in both space and time $\zeta\to\zeta+p^\zeta/2$, $\tau\to\tau+p^\tau/2$. 
Hence the waveform essentially repeats after a distance of half a spatial period, which defines the typical spatial scale of a growth-decay cycle. 
The genus-3 solution in Fig.~\ref{fig:solutions} {\bf b} illustrates that finite-gap solutions can have rather intricate space-time dependence. The individual periods $p^\tau_{ j}=2\pi/\omega_j$ are such that $4p^\tau_1\approx 5p^\tau_2\approx 6p^\tau_3$, and the solution approximately repeats after $\tau=4p^\tau_1\approx 3$. The frequencies are not necessarily commensurate, so that not every finite-gap solution is periodic. In general, finding a spectrum for which the solution has a desired period is a difficult problem~\cite{Belokolos1994}.

\subsection*{The nonlinear wave of the experiment}

We focus on a periodic genus-2 solution for the experiment. The fiber link with periodic amplification can be approximated by an \emph{effective} loss-less form of the integrable NLSE~\eqref{eq:nlse}, provided the typical spatial scale of the solution is large compared to the span length $L$ (see Methods). The equivalent integrable system has the same length as the experimental fiber link, but is governed by a renormalized nonlinearity parameter $\gamma_\text{eff}$.
From~\eqref{eq:thetasol} we obtain a solution to the dimensionful NLSE through the substitutions $z = (T_0^2/|\beta_2|)\zeta$, $T=T_0\tau$ and $A(z,T) = \sqrt{|\beta_2|/(\gamma_\text{eff} T_0)}\psi(\zeta,\tau)$, where $\beta_2$ is the group velocity dispersion, $\gamma_\text{eff}=\gamma L_\text{eff}/L$ is the effective nonlinearity parameter and $L_\text{eff}=[1-\exp(-\alpha L)]/\alpha < L$ the effective interaction length~\cite{Agrawal2000}. 

\begin{figure}[t]
 \center{
\includegraphics[width=0.576\textwidth]{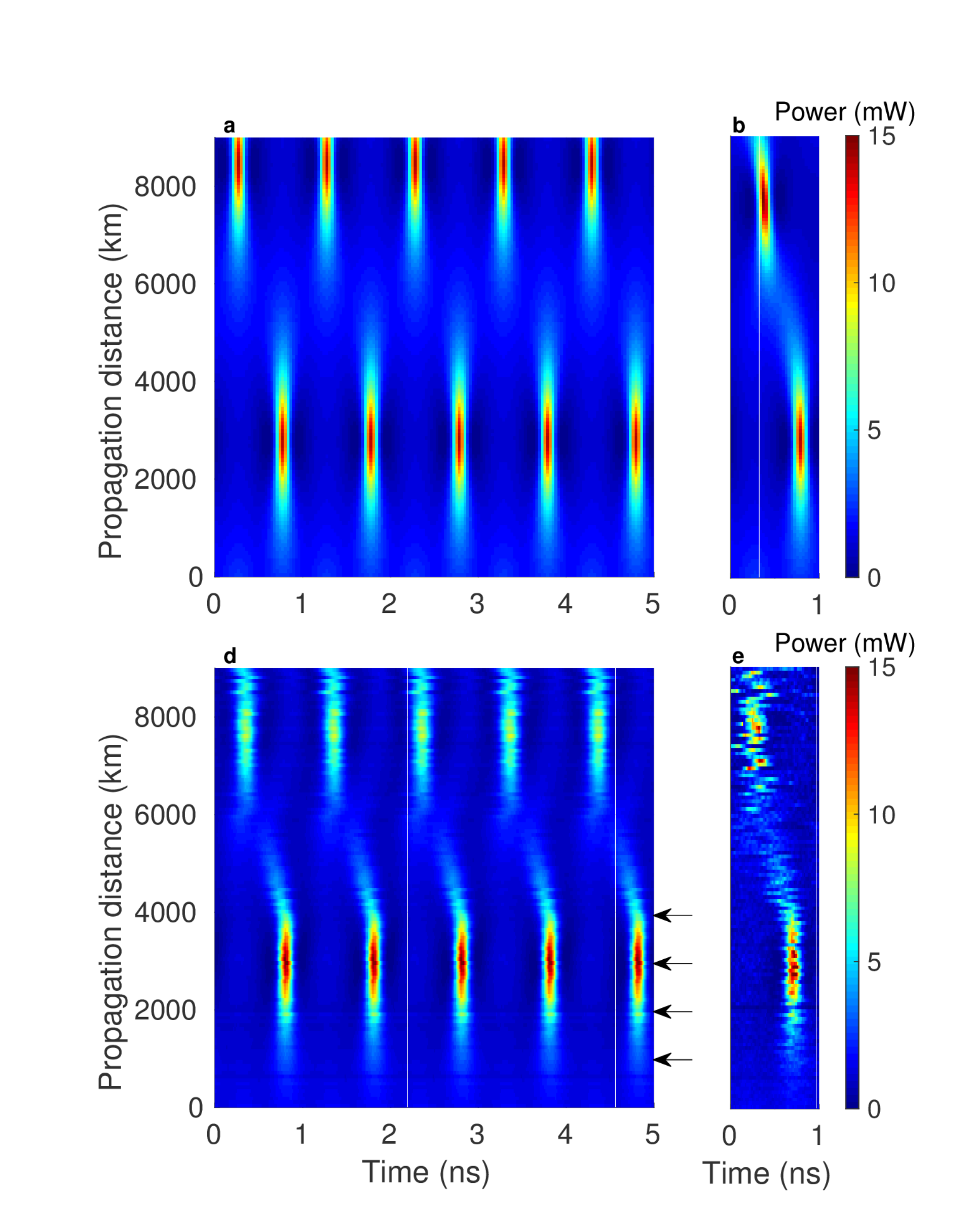}
\includegraphics[width=0.255\textwidth]{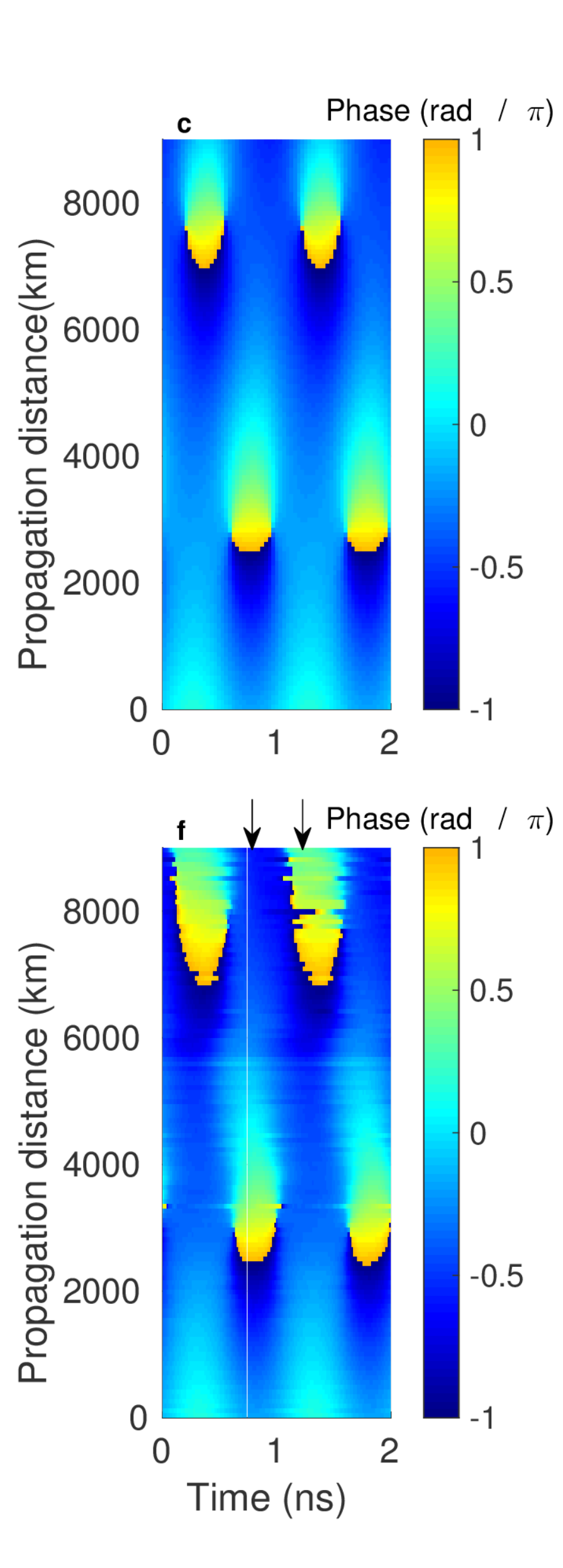}
}
 \caption{\label{fig:surface} {\bf Space-time evolution of genus-2 finite-gap solution}. {\bf a}, {\bf c}, Theoretical prediction of the evolution of the power and phase according to Eq.~\ref{eq:thetasol}. {\bf b}, NLSE simulation including attenuation, periodic amplification and a Gaussian-shaped filter.
{\bf d}, {\bf f}, Experimental data averaged over 25 realizations and after timing correction. {\bf e}, Raw experimental data.
}
\end{figure}

Note that $T_0$ can be chosen freely to set the timescale of the solution. 
We haven chosen $T_0$ to obtain a time period $p^T=1$ ns to sample the waveform with reasonably high accuracy.
In order to guarantee sufficiently slow evolution of the waveform we adjust the spectrum to give a spatial half-period that is a large approximate multiple of the span length $p^z/2 \gg L$.
By further choosing a spectrum with a large imaginary part and closely spaced points, we can obtain a significant peak power compared to the background~\cite{Smirnov2013}.
A suitable spectrum is $\lambda_j = \{-1\pm 4.5i, \pm 5i, 1\pm 4.5i\}$.
The resulting finite gap-solution has a recurrence length (half-period) of $p^z/2=5760$ km ($\sim$77 fiber spans) and average power of $2.1$ mW.
The space-time dependence of the power and phase of the waveform as obtained from Eq.~\ref{eq:thetasol} and scaled to the experimental units are shown in Fig.~\ref{fig:surface} {\bf a} and {\bf c}, respectively. 
The point of maximal temporal compression occurs at 2880 km ($\sim$38 spans) with a peak power of 13.9 mW, roughly six times larger than that of the initial condition. A second peak reaches its maximum after 8640 km.

\subsection*{Measurements}

\begin{figure}[b]
 \center{\includegraphics[width=0.6\textwidth]{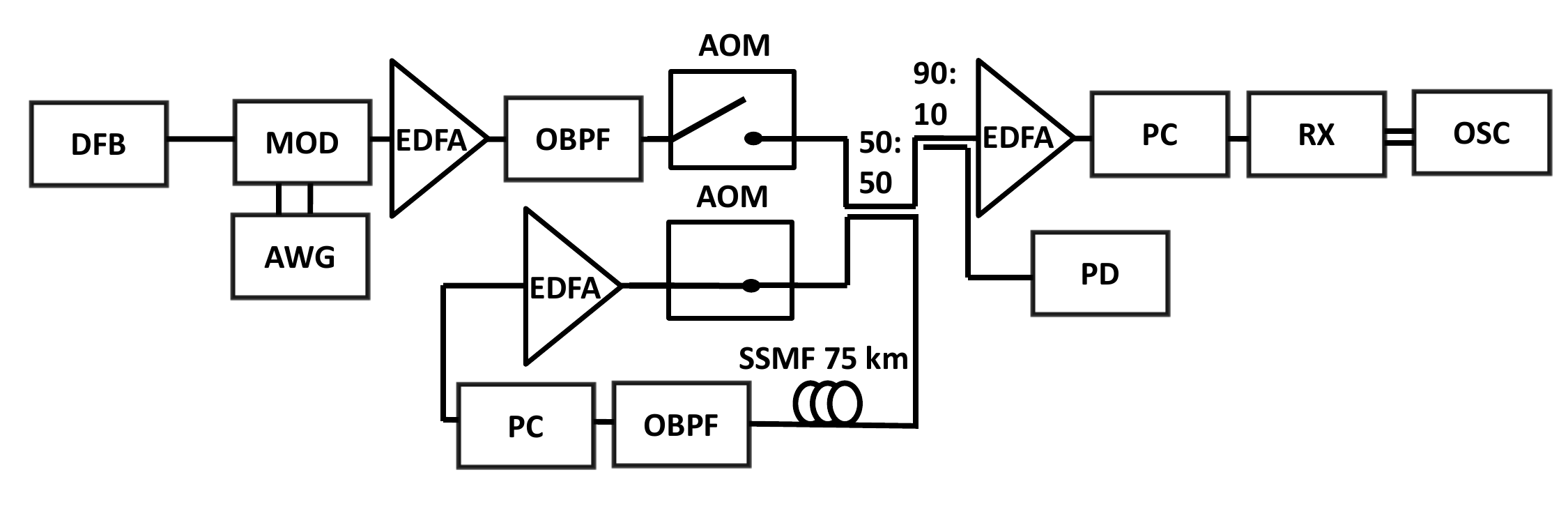}}
 \caption{\label{fig:setup} {\bf Experimental setup.} DFB: distributed feedback fiber laser; MOD: optical modulator, AWG: arbitrary waveform generator; OBPF: optical bandpass filter; EDFA: Erbium-doped fiber amplifier, AOM: acousto-optical modulator; PC: polarization controller; SSMF: standard single mode fiber; PD: photodetector; RX: balanced receiver; OSC: real-time sampling oscilloscope.}
\end{figure}

We performed measurements with the experimental setup shown in Fig.~\ref{fig:setup}.
A telecom-grade narrow linewidth distributed feedback fiber laser (1550 nm) is modulated by a pair of Mach-Zehnder modulators in an interferometric structure relatively phase-shifted by $\pi/2$ to transfer the complex electrical signal to the optical domain. The modulators are driven by an arbitrary waveform generator (AWG) which allows us to generate the initial condition in both amplitude and phase with a high sampling rate of 64 Gs/s per second, or 64 samples per period.
An EDFA amplifies the signal to the desired theoretical average power at the entrance of the fiber.
Losses due to fiber attenuation and components are compensated by the EDFA inside the loop.
A third EDFA after the loop ensures a significant electrical amplitude at the oscilloscope. 
A balanced coherent receiver is used to separate real and imaginary parts of the waveform before it is sampled and digitized by a real-time oscilloscope (80 Gs/s). 
All further data processing, in particular downsampling to 64 Gs/s, timing synchronization, scaling and phase retrieval is done off-line in software (see Methods).

We inject multiple signal bursts consisting of a timing synchronization symbol and 5000 periods of the waveform into the fiber loop (first acousto-optical modulator (AOM) closed and second open to stop propagation of a previously recirculating signal). The number of bursts is determined by the roundtrip time (360$\mu$s) of the signal inside the loop. When the loop is filled, the setting of the AOMs is reversed, blocking the signal coming from the modulator and letting the signal recirculate inside the loop. 
Half of the signal is coupled out of the loop via a 50:50 coupler at each revolution. An electric pulse generator synchronizes the operation of the AOMs and the oscilloscope to record the signal after a preset number of rotations in the loop.
We vary the number of spans between 0 (\emph{back-to-back} configuration) and 120, covering a total  distance of 9000 km, corresponding roughly to the distance between Paris and Los Angeles. Amplitude and phase of the propagating wave are thus recorded with a spatial resolution of 75 km.

\subsection*{Space-time evolution}

Figure~\ref{fig:surface} {\bf d} is a false-color plot of the experimentally measured space-time evolution of the power. It has been obtained after timing synchronization, scaling and averaging over 25 samples of 5 consecutive periods for each distance (see Methods section for details). We can see good agreement with the theoretical waveform shown in Fig.~\ref{fig:surface} {\bf a}. In particular we see the buildup of the first and second peak and surrounding troughs of low intensity. Fig.~\ref{fig:surface} {\bf e} shows a single realization of a single period. Depending on the realization of the amplified spontaneous emission (ASE) noise from repeated signal regeneration, the measurement can deviate from the theoretical prediction. As a result the second peak in Fig.~\ref{fig:surface} {\bf d} is smeared out in the averaging procedure.

The exact solution of Fig.~\ref{fig:surface} {\bf a} has high symmetry (inversion symmetry $\omega\leftrightarrow -\omega$ in frequency), which is apparently broken in the experimental data. In particular one can see a 'tail' emerging from the first peak which crosses over into the second. The troughs can be seen to be located slightly asymmetrically around the peak.
We emphasize that this qualitative deviation is not due to the fact that the system is only approximately integrable due to attenuation, because such deviation cannot break the symmetry.
Deviations from the theoretical prediction occur where temporal compression and associated spectral broadening set in and are due to the tunable optical bandpass filter (OBPF). The Gaussian-shaped OBPF inside the loop has a measured full width at half maximum (FWHM) of 147GHz. Assuming a misalignment of the filter with respect to the center frequency by 5GHz, one obtains a small, but measurable asymmetry in the spectrum. 
The AOMs further cause a small drift of the signal in frequency in each pass through the fiber loop.
Fig.~\ref{fig:surface} {\bf b} shows a simulation based on the NLSE for the effective model, including a Gaussian-shaped filter of measured width, which is assumed to be offset by 5GHz. The result qualitatively reproduces the features observed in the experiment.

\begin{figure}[t]
\center{\includegraphics[width=1.\textwidth]{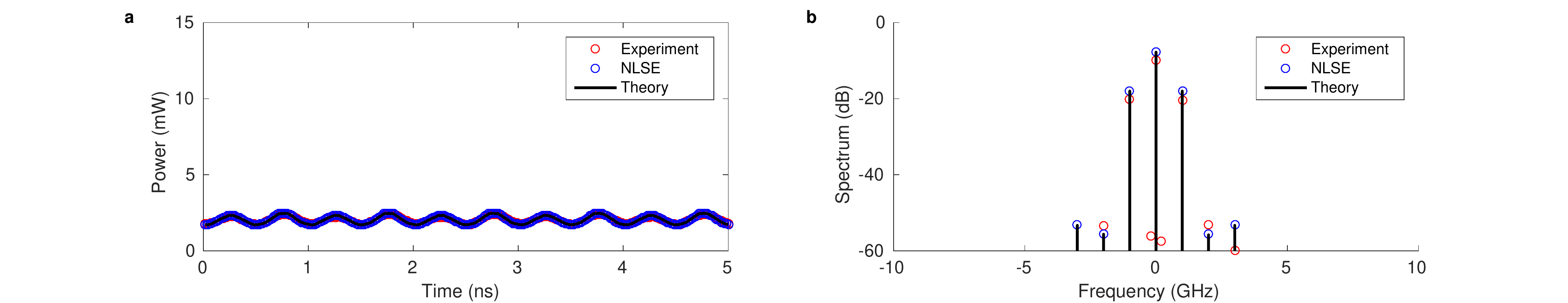}}
\center{\includegraphics[width=1.\textwidth]{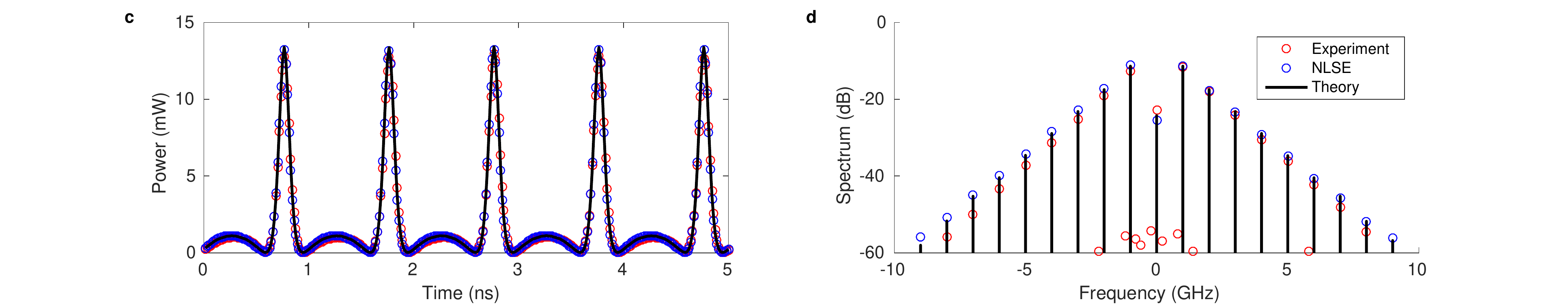}}
\caption{\label{fig:waveform} {\bf Comparison of power and Fourier spectrum between theory, simulation and experiment at fixed distance.} The data corresponds to horizontal cuts in Fig.~\ref{fig:surface}. {\bf a}, Waveform measured in back-to-back configuration compared to the theoretical solution~\eqref{eq:thetasol} and NLSE simulation. {\bf b}, The associated Fourier spectrum.
{\bf c}, Waveform measured after 2925 km (39 spans) at the point of maximal temporal compression. {\bf d}, The Fourier spectrum is taken at the point of maximal spectral broadening and shows the typical triangular shape.
}
\end{figure}

\begin{figure}[b]
\center{
\includegraphics[width=.33\textwidth]{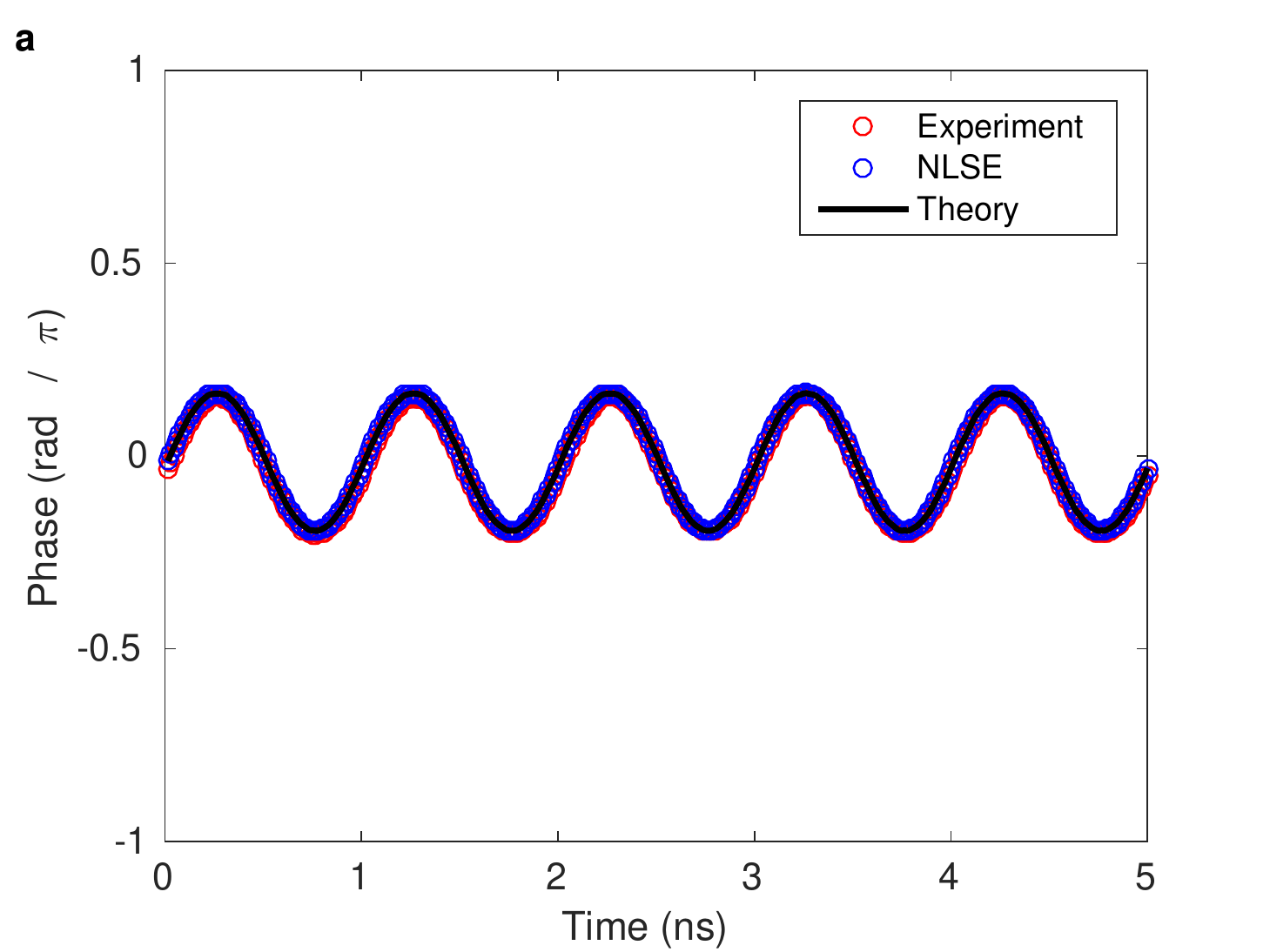}
\includegraphics[width=.33\textwidth]{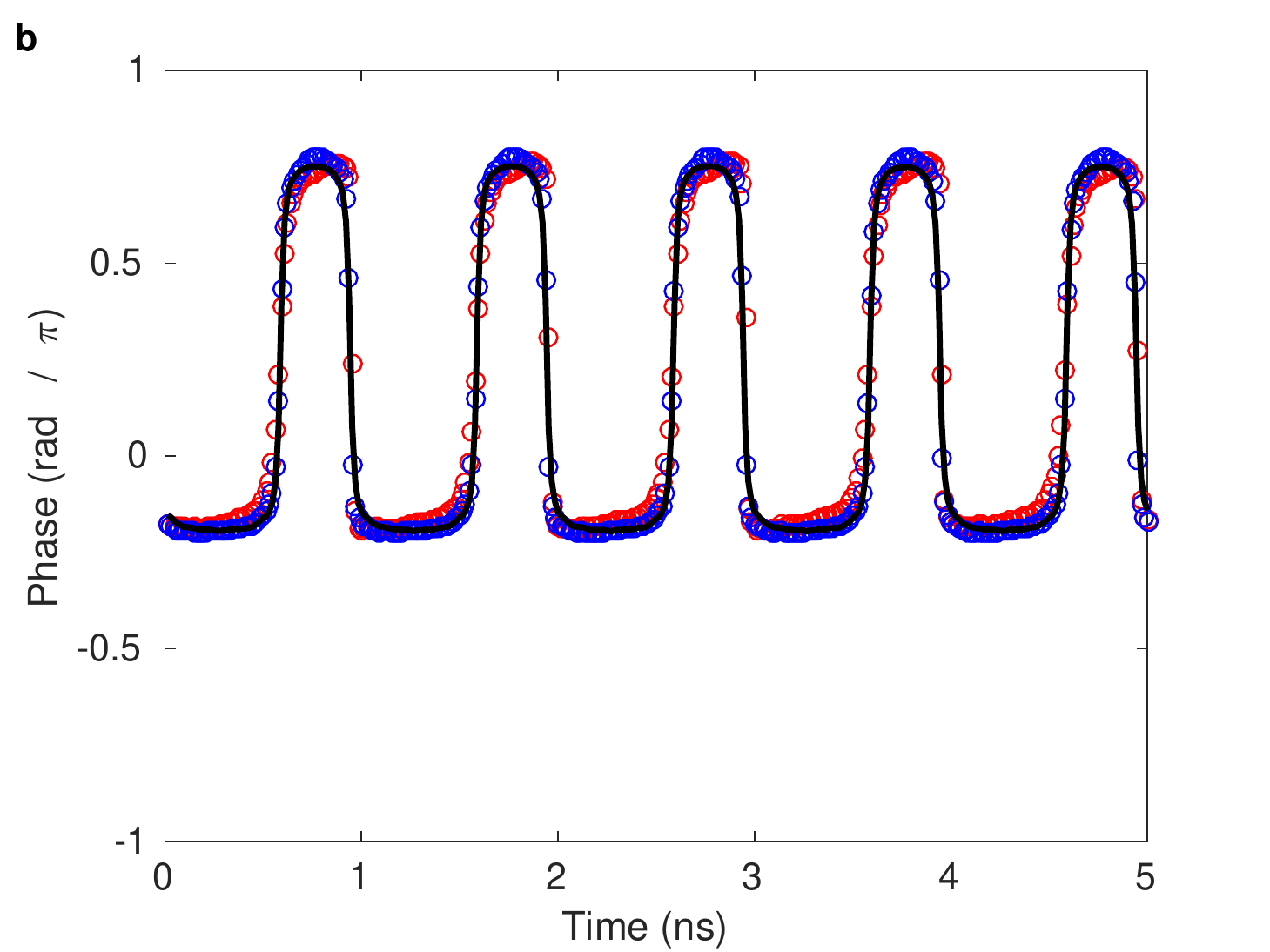}
\includegraphics[width=.33\textwidth]{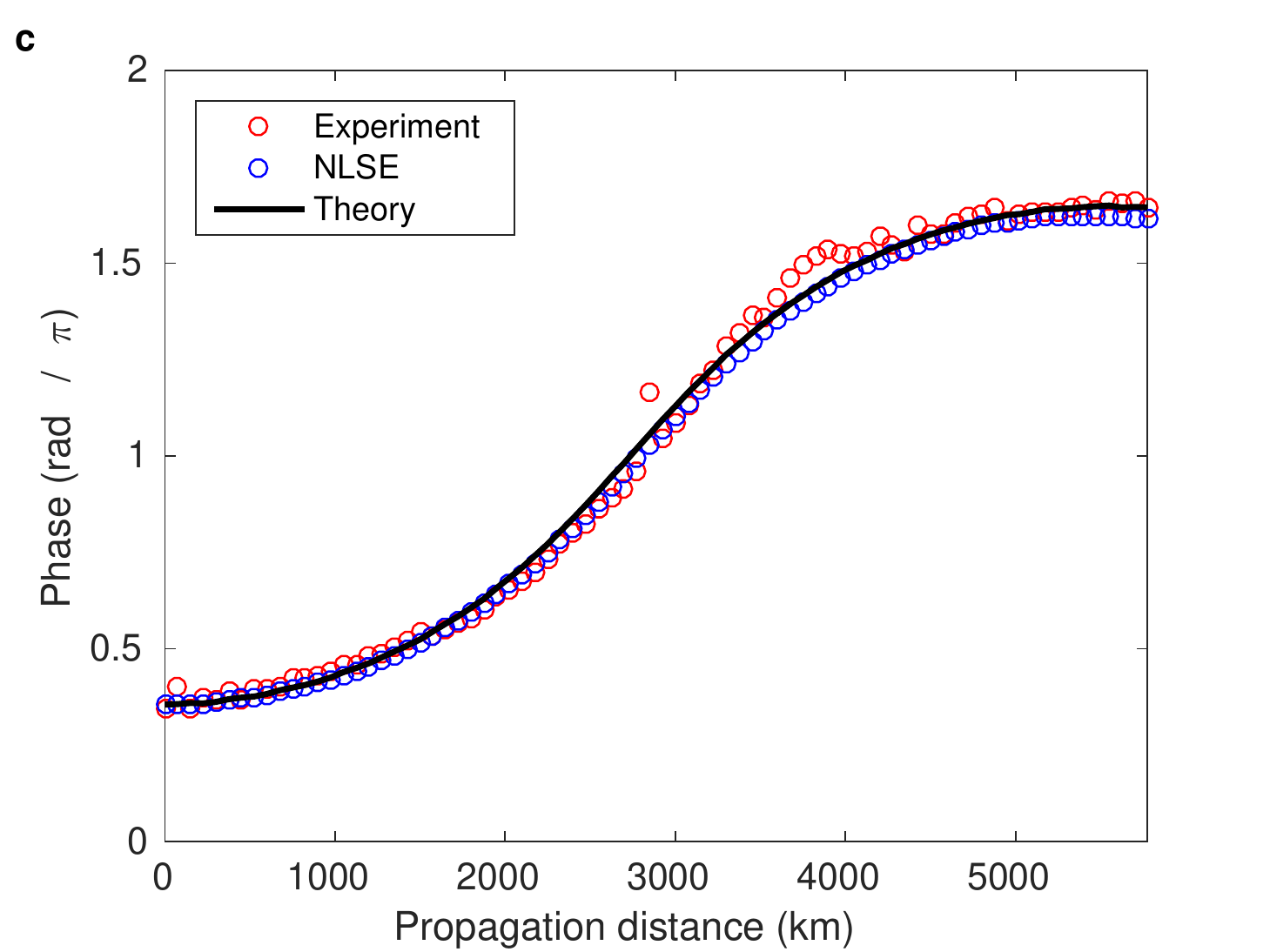}
}
\caption{\label{fig:phase} 
{\bf Comparison of phase between theory, numerical simulation of the NLSE and experiment.} {\bf a}, Phase in back-to-back configuration at fixed distance. {\bf b}, Phase after 2925 km. {\bf c}, Phase difference between two vertical cuts of Fig.~\ref{fig:surface} {\bf f} for fixed times, one passing through the peak and one between two peaks, showing the accumulated phase of the peak relative to its background.
}
\end{figure}

In Figs.~\ref{fig:waveform} and~\ref{fig:phase} we show slices of the experimental data at characteristic distances. 
Figure~\ref{fig:waveform} {\bf a} shows the power of the waveform measured after passing through the setup of Fig.~\ref{fig:setup} without the fiber loop. It illustrates that we can precisely target the initial condition.
99\% of the spectrum shown in {\bf b} is confined in a bandwidth of 2 GHz.
We compare the experimental data with the theoretical result provided by Eq.~\ref{eq:thetasol} scaled to the experimentally measured power and to split-step simulation of the NLSE (see Methods). All NLSE simulations include the effect of fiber attenuation and the optical bandpass filter inside the recirculating loop. Figures~\ref{fig:waveform} {\bf c} and {\bf d} show data at the point of maximal temporal compression in excellent agreement with the theoretical prediction.
The spectrum has broadened by roughly a factor of 5 (99\% of spectrum in 10 GHz bandwidth). The comb decays following a geometric progression with a maximum in the first sideband, showing that the energy is distributed over an infinite number of modes. It leads to the triangular shape that is characteristic of many parametrically driven systems~\cite{Akhmediev2011} and which has been derived analytically for a particular exact periodic solution of the NLSE~\cite{Akhmediev1986}. The triangular comb has been experimentally observed for the Akhmediev breather~\cite{Dudley2009} and Kuznetsov-Ma soliton~\cite{Kibler2012}. After this point the process is reversed, leading to FPUT recurrence. 

Excellent agreement is also observed for the phase data shown in Fig.~\ref{fig:phase} {\bf a} and {\bf b}, reconstructed from averaged measurements of the real and imaginary parts of the waveform. 
The phase difference between the peak and the background at the point of maximal temporal compression shown in {\bf b} is very close to $\pi$, which can also be seen from Fig.~\ref{fig:surface} {\bf c} and {\bf f}.
In Fig.~\ref{fig:phase} {\bf c} we plot the difference between phases on two slices of fixed time, one traversing the peak ($T=0.76$ ns) and one traversing the background ($T=1.26$ ns, see arrows in Fig.~\ref{fig:surface} {\bf f}) as a function of distance. Over a spatial half-period the relative phase difference between the peak and the background accumulates to roughly $1.2\pi$. For the Peregrine soliton a shift of $2\pi$ has been observed~\cite{Xu2019}. Because of the periodicity of the solution we do not observe this limit.

\begin{figure}[t]
 \center{\includegraphics[width=\textwidth]{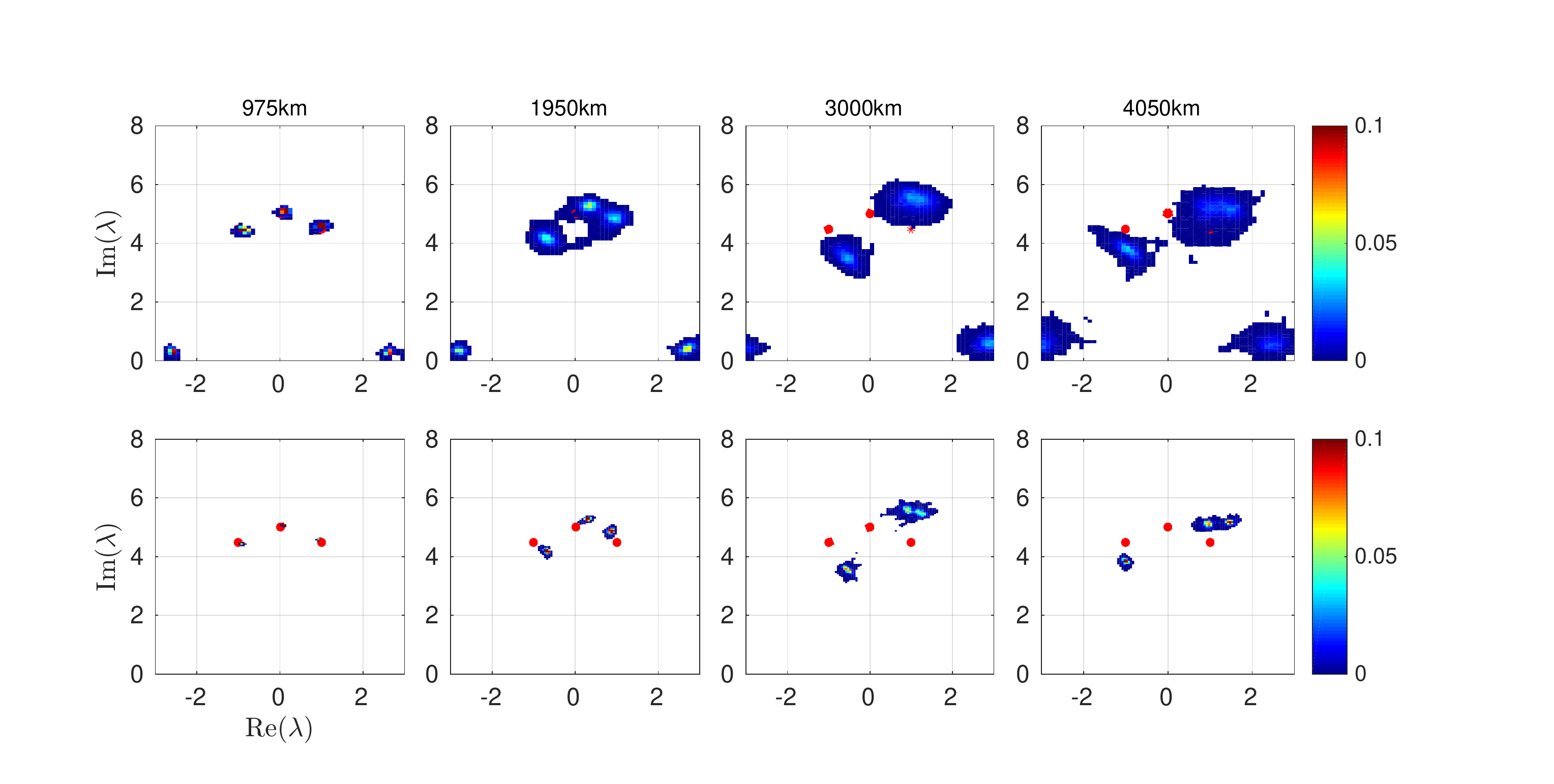}}
 \caption{\label{fig:nlspectrum} {\bf Histograms of nonlinear main spectra at characteristic distances.} The nonlinear main spectra computed based on the experimental raw data for the waveform (top) and using the averaged data (bottom). Noise accumulates with distance due to repeated amplification, leading to distributions around the theoretical prediction (red dots) and artifacts with small imaginary part along the real axis. In the spectra obtained from the averaged data, the three distinct spectral points of the genus-2 finite-gap solution can be distinguished for all distances. Note that while for 3000 and 4050 km the point clouds have merged, they contain two discernible peaks of high intensity. Departure from the theoretical prediction is due to the optical filter, see text.}
\end{figure}

\subsection*{Nonlinear spectral analysis}

We have shown that the waveform qualitatively evolves according to the theoretical prediction. 
In order to unambiguously demonstrate that the propagating wave is a finite-gap solution, we perform nonlinear spectral analysis of the real and imaginary components of the received data.

Nonlinear spectral analysis has previously been used to characterize SFB solutions and to identify similar structures in noise seeded MI~\cite{Randoux2016,Randoux2018}. 
Since SFB solutions correspond to degenerate cases of finite-gap solutions with an infinite period, the relevant peak structures have to be cut and periodized to obtain a non-degenerate spectrum, which, as a consequence, is not exactly constant even if the dynamics were ideally integrable. In our case we would expect to exactly recover the main spectrum of Fig.~\ref{fig:nlspectrum} if the dynamics are integrable (and without spectrum distortions). In NLSE simulations of the non-integrable system including attenuation we observe small oscillations of the main spectrum with distance, which however preserves its original shape and symmetry as expected.

We compute the nonlinear spectrum from the waveform via a fast nonlinear Fourier transform algorithm~\cite{Wahls2015} implemented in the open-source FNFT package~\cite{Wahls2018}. 
The algorithm is applied to one period of the raw data at a time. The resulting complex spectrum values are collected in a histogram. To obtain sufficient statistics we use five bursts of 5000 periods. 
The result is shown in the top row of Fig.~\ref{fig:nlspectrum} at a few characteristic distances (see arrows in Fig.~\ref{fig:surface} {\bf d}).
At 975 km (15 spans) one can clearly distinguish three clouds of spectral points centered around the theoretical values as expected for a genus-2 finite-gap solution. The spread of the spectrum values is due to noise which primarily stems from the ASE of the optical amplifiers. Additional eigenvalues at small imaginary part arise because the periodic waveform affected by noise deviates from a pure genus-2 solution. Further similar artifacts occur along the real axis (not shown).
As the distance increases, the noise distributions grow as ASE noise is added due to repeated amplification. While three eigenvalues can still be distinguished at 1950 km (26 spans), the distributions have merged after 3000 km (40 spans).

In the bottom row of Fig.~\ref{fig:nlspectrum} we show data obtained by applying the algorithm to averages over 25 periods to reduce the noise. We note that the procedures of averaging and nonlinear Fourier transform approximately commute, suggesting that noise is a small perturbation. The noise in the resulting spectra is considerably reduced.
After 975 km (and in back-to-back configuration, not shown) the result almost perfectly agrees with the theoretical spectrum, which shows that departure from integrability is a small effect. 
Deviations of the spectrum values from the theoretical prediction start to occur around 1950 km, where spectral broadening sets in and the optical filter distorts the spectrum. Our NLSE simulations qualitatively reproduce the observed deviations from the ideal nonlinear spectrum as a result of the skewing of the Fourier spectrum through the optical filter.
At 3000 km and beyond, the point clouds corresponding to two of the three spectral points have merged, but the distributions retain two discernible peaks of high intensity in each case. These results unambiguously confirm the propagation of a genus-2 finite-gap solution in our experiment.

\section*{Discussion}

We have observed the space-time evolution of a periodic finite-gap solution of the integrable NLSE in close agreement with theoretical predictions. 
The experiment has been conducted in a testbed for optical transmission experiments built from commercially available standard components.
Modern coherent telecommunication equipment is designed for modulation and detection of both amplitude and phase with high sampling rates.
Hence we were able to precisely target the initial condition and measure the real and imaginary parts of the complex envelope of the electrical field with high accuracy.
Together with the recirculation loop, the setup permits a non-destructive full-field characterization of the spacetime dependence of the propagating wave. 
This allowed us to experimentally verify a key prediction of the finite-gap integration theory: through nonlinear spectral analysis we showed that the nonlinear main spectrum is preserved over transcontinental distances. The prospect of encoding information in the invariant nonlinear spectrum is of high interest for applications in optical communication.

We further found that despite the presence of ASE noise from repeated amplification, the evolution of the optical field is accurately described by the integrable NLSE. Even though ASE noise becomes a limiting factor at large distances, the remarkable agreement between experiment and theory shows that such a setup provides highly controlled conditions for observing nonlinear wave dynamics. 

The observed periodic compression dynamics constitute a realization of FPUT recurrence, which we have observed for close to two growth decay cycles.
The results are in line with recent observations of the space-time dependence of FPUT recurrence~\cite{Kimmoun2016,Pierangeli2018,Mussot2018,Naveau2019}.
In particular, our results bear remarkable qualitative resemblance to the very recent ones of Ref.~\cite{Naveau2019}, while exceeding them in terms of scale in space and time by three orders of magnitude. At these scales the NLSE is an excellent model of the nonlinear dynamics. In their experiment, loss is compensated by a Raman pump. A small number of waves (pump, signal, idler and harmonics) is launched into the fiber and the time-domain signal is reconstructed by means of the Fourier transform of up to five waves. The propagating and reconstructed waves therefore correspond to time-periodic fields similar to the one in our experiment. Here we even achieve close quantitative agreement with the underlying theory.

We have observed FPUT recurrence over more than one cycle. Previously, approximate recurrence has been observed over three complete cycles, albeit in a photorefractive KLTN crystal~\cite{Pierangeli2018}. In order to observe multiple recurrence cycles with our setup, the number of spans can be increased, or the spatial period of the waveform may be reduced for fixed fiber length. Both lead to deviations from integrability: due to increasing noise for the first option and a departure from the validity of the effective model for the second. For the latter, the effect can be mitigated in two ways. By either decreasing the span length, again decreasing the signal to noise-ratio, or by increasing the separation between nonlinear spectrum points, which reduces the dynamic range of the signal power. Distributed Raman amplification has the potential to provide a flat gain to further increase the achievable distances.

While our experiment has been conducted neglecting polarization effects, the two polarization components of single-mode fibers and the various modes in multi-mode fibers are described by the integrable Manakov system~\cite{Wai1996,Mumtaz2012}. This opens possibilities to design analogous experiments on generalizations of finite-gap solutions to the Manakov system~\cite{Christiansen2000} or for the study of vector solitons.

\section*{Methods}

\subsection*{Inverse scattering transform and finite-gap integration}

The IST is a well-established method to solve certain classes of integrable equations. It was first discovered for the Korteweg-de Vries equation~\cite{Gardner1967} and later by Zhakharov and Shabat~\cite{Shabat1972}for the NLSE.
In the IST, the NLSE is represented as the compatibility condition $\partial_\tau \partial_\zeta \Phi = \partial_\zeta \partial_\tau \Phi$ of two \emph{linear} differential equations for the scattering data $\Phi(\zeta,\tau;\lambda)$,
\begin{equation}
\frac{\partial}{\partial \tau}\Phi = \left[-i\lambda\sigma_3 + \left(\begin{array}{cc}0 & -\psi\\ \bar{\psi} &0\end{array}\right)\right]\Phi \equiv \mathcal{U}\Phi,\qquad \frac{\partial}{\partial \zeta}\Phi = 
\left[-2\lambda\mathcal{U} + \left(\begin{array}{cc}
-i|\psi|^2 & i\frac{\partial\psi}{\partial\tau}\\i\frac{\partial\bar{\psi}}{\partial\tau} &i|\psi|^2 
\end{array}\right)\right]\Phi \equiv \mathcal{V}\Phi,
\end{equation}
which holds true if and only if $\psi(\zeta,\tau)$ satisfies the NLSE. For given initial condition $\psi(\zeta=0,\tau)$, obtaining the scattering solution $\Phi(\zeta=0,\tau;\lambda)$ is known as the Zakharov-Shabat scattering problem. 
The spatial evolution of the scattering data is governed by $\mathcal{V}$ and allows one to determine the solution of the NLSE at any $\zeta$ by solving the inverse scattering problem, that is, by recovering the potential from the scattering data.
The main spectrum of a periodic finite-gap solution is defined by the $2N$ discrete values $\lambda_j$ which occur in complex conjugate pairs and for which the scattering solution,  i.e. the Bloch functions, are (anti-)periodic in $\tau$: $\Phi(\zeta,\tau;\lambda)=\pm \Phi(\zeta,\tau+T;\lambda)$. 

Solutions of the NLSE can be constructed by solving a Riemann-Hilbert problem~\cite{Kotlyarov2017}, or the finite-gap integration method, an algebro-geometric approach~\cite{Belokolos1994,Tracy1988}. Here we adopt the latter method to construct a periodic solution~\eqref{eq:thetasol} by computing its parameters in terms of integrals over the hyperelliptic Riemann surface defined by the main spectrum.
To this end, we define a canonical homology basis $\{a_i, b_j|1\leq i,j\leq g\}$  of $a$- and $b$-cycles on the two-sheeted Riemann surface such that each $a$-cycle crosses one $b$-cycle exactly once and otherwise crosses no other $a$- or $b$-cycle, and the same holds for the $b$-cycles. This is expressed in terms of the intersection numbers as $a_i\circ b_j = \delta_{i,j}$, $a_i\circ a_j = b_i\circ b_j = 0$. The procedure can be automatized and  for example yields the homology basis shown in Fig.~\ref{fig:cycles} {\bf a}.
In terms of the basis of holomorphic differentials $dU_j = \lambda^{j-1} d\lambda/\sqrt{P(\lambda)}$, the invertible $g\times g$ matrices of $a$- and $b$-periods are defined in terms of the integrals $\bm{A}_{jk} = \oint_{a_k}dU_j$, and $\bm{B}_{jk} = \oint_{b_k}dU_j$. 
We evaluate these integrals numerically.
From these we obtain the frequencies  $\omega_j = -4\pi i (\bm{A}^{-1})_{j,g}$, wavenumbers $k_j = -8\pi i ((\bm{A}^{-1})_{j,g-1} + \sum_{k=1}^{g+1}\mathrm{Re}\,{\lambda_k}\,(\bm{A}^{-1})_{j,g})$ and the period matrix $\bm{\tau} = \bm{A}^{-1}\bm{B}$.
In terms of the normalized differentials $d\underline{V} = \bm{A}^{-1}d\underline{U}$ the phases $\delta_j^{\pm}$ are given by
\begin{equation}
\frac{1}{2\pi}\delta^\pm_j = \int_{p_0}^{\infty^\pm}dV_j-\frac{1}{2}\tau_{jj} - \sum_{m=1}^g\left(\int_{p_0}^{\mu_m(0,0)}dV_j - \oint_{a_m}V_j dV_m \right).
\end{equation}
They involve the initial condition for the auxiliary variables.
However their difference $\underline{\delta}^+ - \underline{\delta}^- = \int_{\infty^-}^{\infty^+} d\underline{V}$ is independent of $\mu_j(0,0)$.
Not every initial condition of the $\mu_j$ corresponds to a finite-gap solution. 
Since we are not interested in the precise evolution of the $\mu_j$, we exploit the fact that any choice of $\underline{\delta}^-$ yields a valid solution as long as its imaginary part is zero~\cite{Tracy1988}. Hence we can set $\delta_j^-\equiv 0$ and compute $\delta_j^+$ using the formula for their difference.
For a finite-gap solution half of the $2N$ real degrees of freedom in the $\mu_j$ are constrained so that this choice completely determines the auxiliary spectrum.
The remaining parameters $K$, $k_0$ and $\omega_0$ are determined\cite{Belokolos1994} from integrals of normalized Abelian differentials of the second and third kind to $\infty^\pm$. The integration path is chosen such that it crosses a branch cut once to change sheets, while avoiding the holomorphic basis of differentials. see Fig.~\ref{fig:cycles} {\bf a}.
The Riemann $\theta$-functions finally are evaluated numerically for given argument by truncating the series according to a precision criterion.

\subsection*{Experiment and data preprocessing}		

To map the spatiotemporal evolution of the waveform we perform independent measurements for each distance (number of turns in the recirculating loop).
The AWG periodically generates sequences of 5000 periods of periodic waveform (5$ \mu$s) separated by a 42 ns guard interval and preceded by a Schmidl-Cox synchronization symbol~\cite{Schmidl1997} (42 ns) separated from the waveform by a 21ns guard interval.
We employ a telecommunication-grade distributed feedback fiber laser with linewidth $\sim 100$ Hz operating at 1550 nm. 
The optical IQ modulator (Fujitsu 32GB DP-QPSK) is driven by an arbitrary waveform generator with a sampling rate of 64 Gs/s, bandwidth 25 GHz, and effective number of bits (ENOB) of 5.5. Out-of-band spectral components from the modulator are filtered using a rectangular optical bandpass filter (OBPF, Yenista XTA-50). To remove any DC components of the envelope (the low-frequency limit of electrical drivers of the MZ is 35 kHz), we digitally shift the spectrum by $f=5$ GHz.
Insertion losses of the acousto-optic modulators (AOM) and couplers are compensated by an Erbium-doped Fiber amplifier (EDFA).
At system setup we open the recirculating loop and adjust the power setting of the first EDFA to obtain the desired theoretical average power at the entrance of the fiber.
With closed loop and circulating signal, the EDFA inside the loop is adjusted based on the output of the low-bandwidth photodetector (PD) signal to compensate fiber attenuation and losses due to components.
The recirculating fiber loop itself consists of 75km of standard single-mode fiber (G.652) and a Gaussian-shaped OBPF (measured FWHM 147GHz) to suppress out-of-band noise. 
We minimize power fluctuations in the loop due to polarization dependent loss through a polarization controller (PC). The power is monitored by a low-bandwidth photodetector.
An electronic pulse generator is used to control the number of roundtrips inside the loop through the AOMs and to synchronize the oscilloscope. 
A second PC gathers the signal in a single polarization component.
The output of the intradyne phase diversity coherent receiver is proportional to the amplitude of the in-phase and quadrature components of the signal. They are independently digitized using an 80Gs/s oscilloscope (Keysight DSOZ504A, ENOB 6 per sample).
Each pass through an AOM downshifts the center frequency by 40 MHz. On the receiver side, the 5 GHz frequency shift and any residual carrier phase offset caused by the AOMs are compensated by shifting the center frequency of the spectrum to 0 Ghz (in digital domain). The signal exhibits a peak at multiples of 1 GHz due to its periodicity, which can be used as a reference.
The signal in digital domain is further downsampled to 64Gs/s. An approximate relative timing to locate the relevant data is obtained through the Schmidl-Cox synchronization algorithm. Due to noise in the transmission, the timing can be off by up to 6 samples~\cite{Schmidl1997} (64 samples per period).
The data are processed independently for each distance. To avoid edge effects, 100 periods on each end are discarded. The remaining 4800 periods are subdivided into groups of 25$\times$5 periods. For each group of the 40 groups we average over the 25 realizations per group to obtain averaged waveforms of 5 periods length.
Since our setup only detects relative phase, absolute phase for each distance is recovered by aligning the phase at the maximum amplitude at the respective distance to the corresponding theoretical value.
Sample-accurate timing synchronization is finally achieved by aligning the phase of the averaged data (within a window of one period) to the phase of the theoretical waveform at a fixed time (T=4.875 ns).
The same timing correction is applied to real and imaginary parts of the averaged and unaveraged data.

\subsection*{Simulations and effective model}

The dimensional NLSE describing nonlinear space-time evolution of the complex envelope $A(z,T)$ of the electric field of the optical carrier is given by~\cite{Agrawal2000}
\begin{equation}
i\frac{\partial A(z,T)}{\partial z} - \frac{\beta_2}{2}\frac{\partial^2A(z,T)}{\partial T^2} = - \gamma|A(z,T)|^2A(z,T) -\frac{i\alpha}{2}A(z,T),
\label{eq:nlsedim}
\end{equation}
where $z$ denotes distance, $T$ is time a reference frame moving with the group velocity of the envelope. At the scale of our experiment, terms describing third-order dispersion, self-steepening and the delayed Raman response can be neglected. Despite the use of low-loss telecommunication fibers attenuation becomes significant and renders the system non-integrable. Substituting $A(z,T)=\exp(\frac{-\alpha z}{2})\tilde{A}(z,T)$ the right-hand side transforms into $- \gamma(z)|\tilde{A}(z,T)|^2\tilde{A}(z,T)$, where $\gamma(z)=\gamma\exp(-\alpha z)$.
In terms of dispersion and nonlinear operators the NLSE becomes $\partial A/\partial z = -i[\mathcal{D}+\mathcal{N}(z)]A(z,T)$ and its solution can formally be written
$
A(z,T) = \exp\left(-i\int_0^z dz'[\mathcal{D}+\mathcal{N}(z)]\right)A(0,T).
$
Under the assumption that $\tilde{A}(z,T)$ is approximately constant over a fiber span of length $L$, 
$$
\int_0^L dz' \mathcal{N}(z') = -\gamma\int_0^L dz' |\tilde{A}(z',T)|^2\exp(-\alpha z') \approx -\gamma_\text{eff}|\tilde{A}(z,T)|^2 L,
$$
we obtain an approximately integrable model with renormalized effective nonlinearity parameter $\gamma_\text{eff} = \gamma L_\text{eff}/L$ and $L_\text{eff}=[1-\exp(-\alpha L)]/\alpha < L$ being the effective interaction length~\cite{Agrawal2000}. The above assumption is justified if the amplitude $\tilde{A}$ in the effective model varies sufficiently slowly in space over a fiber span.

From a Gaussian fit of a single pulse (half time-period) of the initial condition (Fig~\ref{fig:waveform} {\bf a}) we obtain a width at $1/e$ height of $T_\sigma=0.28$ ns and a peak power of $P_0=2.4$ mW of the initial condition. We can estimate the nonlinear length $L_\text{NL} = (\gamma_\text{eff}P_0)^{-1}=1150$ km and dispersion length $L_\text{D}=T_\sigma^2/|\beta_2|=3640$ km. Both are significantly larger than the span length, but are shorter than the fiber link. The scales are of the same order of magnitude and cannot be separated. Dispersion and nonlinearity act together in governing the evolution of the waveform.

We perform simulations of the NLSE~\eqref{eq:nlsedim} using a standard split-step integration scheme~\cite{Agrawal2000} with adaptive step size controlled by a maximum phase change of 0.05$^\circ$.
Since the implementation is based on FFT and implicitly assumes periodic boundary conditions, it is sufficient to simulate a single period.
To quantify the departure from integrability, we performed NLSE simulations of the fiber link including attenuation.
We simulate the system consisting of a fixed number of identical fiber spans including the Gaussian-shaped filter with measured FWHM and with lumped amplification. We neglect ASE noise to emphasize small deviations from integrability arising from attenuation and the filter. We further assume a chromatic dispersion coefficient $\beta_2= -21.5\textrm{ps}^2\textrm{km}^{-1}$, nonlinearity parameter $\gamma = 1.3\cdot\textrm{W}^{-1}\textrm{km}^{-1}$ and attenuation $\alpha=0.2$dB/km ($L_{\text{eff}}=21$ km).
The simulations show a relative deviation from the theoretical result for the integrable system of less than 1\% on average, with a maximum difference of 9\% reached at the smallest amplitude.
Idealized NLSE simulations of the effective model neglecting attenuation yield essentially the same results as the theoretical prediction by Eq.~\eqref{eq:thetasol}, confirming the high accuracy of the split-step Fourier integration method.

\bibliography{finitegap}

\section*{Author contributions statement}

H.H. and J.-W.G. conceived the experiment, J.-W.G. and Y.J. conducted the experiment, J.-W.G. and H.H. analyzed the results and wrote the paper. All authors reviewed the manuscript.

\section*{Competing interests}
The authors declare no competing interests.

\end{document}